\documentclass[manuscript,nonacm]{acmart}

\usepackage{multirow,tabularx}
\usepackage{makecell}
\usepackage{float}
\usepackage{subfig}
\usepackage[font=small]{caption}
\usepackage{fontawesome5}
\usepackage{blindtext}
\usepackage{xpatch}
\usepackage{enumitem}

\usepackage{titlesec}

\usepackage{setspace}
\renewcommand\footnotetextcopyrightpermission[1]{} 
\acmJournal{POMACS}
\acmVolume{37}
\acmNumber{4}
\acmArticle{111}
\acmMonth{8}

\begin{document}

\title{Graphing else matters: exploiting aspect opinions and ratings in explainable graph-based recommendations}

\author{Iván Cantador}
\affiliation{
\institution{Universidad Autónoma de Madrid}
\country{Spain}
}

\author{Andrés Carvallo}
\affiliation{
\institution{Pontificia Universidad Católica de Chile}
\country{Chile}
}

\author{Fernando Diez}
\affiliation{
\institution{Universidad Autónoma de Madrid}
\country{Spain}
}

\author{Denis Parra}
\affiliation{
\institution{Pontificia Universidad Católica de Chile}
\country{Chile}
}

\begin{abstract} 
The success of neural network embeddings has entailed a renewed interest in using knowledge graphs for a wide variety of machine learning and information retrieval tasks. In particular, current recommendation methods based on graph embeddings have shown state-of-the-art performance. These methods commonly encode latent rating patterns and content features. Different from previous work, in this paper, we propose to exploit embeddings extracted from graphs that combine information from ratings and aspect-based opinions expressed in textual reviews. We then adapt and evaluate state-of-the-art graph embedding techniques over graphs generated from Amazon and Yelp reviews on six domains, outperforming baseline recommenders. Our approach has the advantage of providing explanations which leverage aspect-based opinions given by users about recommended items. Furthermore, we also provide examples of the applicability of recommendations utilizing aspect opinions as explanations in a visualization dashboard, which allows obtaining information about the most and least liked aspects of similar users obtained from the embeddings of an input graph.
\end{abstract}



\maketitle

\section{Introduction}

In recent years, knowledge graphs (KG) have been widely used in recommendation~\cite{he2015trirank, wang2018ripplenet}. A KG is a graph data structure containing information about semantic entities (or concepts) expressed as nodes, and semantic relations between entities expressed as edges. The relations can be represented as \textit{<subject, property, object>} triples whose elements may belong to structured knowledge bases or ontologies. Subject and object entities (nodes) can refer to users, items, item metadata, and content features, among others.
Compared to recommendation approaches that treat relations between users and items independently, exploiting a KG graph allows capturing deeper relations valuable to generate better recommendations~\cite{wang2019kgat}. Another benefit of a KG-based recommender is its capacity to deal with cold-start situations for new users or unrated items by encoding sequential dependency among them~\cite{hu2018leveraging, huang2018improving}. Moreover, they enable the generation of human-interpretable recommendations, making the systems more reliable and transparent for the end-user ~\cite{wang2019explainable, he2015trirank, xian2019reinforcement}.

The broad majority of graph-based recommendation methods makes use of information underlying rating relations and connections between items with similar attributes~\cite{wang2019explainable}. In general, these methods encode latent rating patterns and content features using graph embeddings, since the information stored in KG triples can be too computationally expensive to manipulate. Knowledge graph embeddings~\cite{cai2018comprehensive} consist of learned lower dimension, continuous vectors for entities and relations that preserve the original structure of the input KG.

In this context, a source of information that has been barely explored for KG is the textual content of user reviews. In the research literature, it has been shown that both ratings and reviews can give support and trust to users in decision making tasks~\cite{lackermair2013importance}. In particular, approaches have been proposed to automatically extract personal opinions about item aspects (i.e., features, components and functionalities) from reviews~\cite{poria2016aspect}, and aspect-based opinions have been used to enrich sentiment analysis models~\cite{de2018aggregated}, document classifiers~\cite{ostendorff2020aspect}, and recommender systems~\cite{he2015trirank}. Concerning the exploitation of aspect-based opinions by recommendation methods, Musto et al.~\cite{musto2019combining} made use of aspects to justify recommendations using sentiment analysis and text summarization, whereas Wu and Ester~\cite{wu2015flame} proposed a probabilistic approach that combines aspect-based opinion mining and collaborative filtering. 

Aspect-based opinions have also been included as additional features for latent factor~\cite{qiu2016aspect} and neural network~\cite{chin2018anr, guan2019attentive} models. To the best of our knowledge, the approach by He et al.~\cite{he2015trirank} is the only one that takes advantage of a KG with information from aspect-based opinions extracted from reviews, by enriching user-item binary relation to an heterogeneous tripartite graph composed of user-item-aspect ternary relations.

Unlike previous approaches, our method combines rating and aspect-based opinion interactions in a KG embedding framework to obtain a richer representation of users, items and aspects in the same latent space for making recommendations. Furthermore, we also provide a use case of the applicability of recommendations using aspect opinions as explanations in a visualization dashboard that allows exploring information about the most and least liked aspects of similar users obtained from the embeddings of an input graph.

Therefore, the main contributions of our work are as follows:
\begin{enumerate}
\item Presenting a graph model that represents and relates users, items and aspects in the same latent space for its exploitation by recommender systems.
\item Using the graph model to adapt recommendation methods based on neural network graph embeddings, showing empirical performance improvements over state-of-the-art baselines. 
\item Proposing a method that uses the graph model to provide explanations of generated recommendations at the aspect opinion level.
\end{enumerate}


\section{Related work}

In this section, we survey relevant literature on four research lines: graph-based recommendations (subsection 2.1), review and text-based recommendations (subsection 2.2), aspect-based recommendations (subsection 2.3), and graph-based recommendation explanations (subsection 2.4).

\subsection{Graph-based recommendations} We can distinguish between three main types of approaches that use graphs for recommendation purposes: path-based, embedding-based, and unified. 
\textit{Path-based methods} take advantage of users and items connectivity score functions to generate recommendations. For instance, Luo et al.~\cite{luo2014hete} proposed Hete-CF, which makes use of path similarity combined with common paths between two entities in a social network to make recommendations. Concerning \textit{embedding-based methods}, a KG is used to obtain latent representations of users and items, with which finding closest users and items. Examples of this type of methods are the approach by Wang et al.~\cite{wang2018dkn} for news recommendation and Huang et al.~\cite{huang2018improving}, which incorporate sequential information through a recurrent neural network, and the recent KGAT~\cite{wang2019kgat} method, which models high order relations in a KG. Finally, \textit{unified methods} exploit information from embedding- and path-based forms together. Wang et al.~\cite{wang2018ripplenet} proposed RippleNet, which introduces the concept of preference propagation to refine user and item representations obtained with embedding-based methods.

In our work, we advocate for embedding-based graph recommendations, since they allow better addressing the cold start problem and efficiently representing entities and relations, preserving the graph structure~\cite{wang2019kgat, wang2018ripplenet}. We also chose the embedding-based approach since the state-of-the-art KGAT method~\cite{wang2019kgat} achieved better performance results than other embedding-based, path-based and unified approaches. This led us to empirically compare our method with KGAT in the experimental section.

\subsection{Review and text-based recommendations}
Several works have used textual information extracted from reviews to improve the modeling of user preferences and recommendation performance. Shet et al.~\cite{shalom2019generative} proposed to use review text as input for a generative recommendation model. Similarly, Chuang et al. proposed the TPR model~\cite{chuang2020tpr}, which makes use of user-item interaction combined with relations between item and associated text for improving the modeling of user preferences. 

Another usage of user-generated content is done by DAML~\cite{liu2019daml}, a dual attention model that learns from ratings and reviews texts inputs. RNS, a review neural recommender model proposed by Li et al. ~\cite{li2019review}, uses attention mechanisms over the users' historical purchases and past reviews for collaborative filtering. Lastly, Wang et al. ~\cite{wang2021leveraging} proposed using other properties extracted from reviews different from text, such as user age and review length, for learning user preferences and item attributes. 

A major benefit of including textual data into a recommendation model is the capability of providing human-readable explanations. One example of this is the work done by Zhang et al.~\cite{zhang2014explicit}, which incorporates explainable sentence-level sentiment information extracted from reviews. In the same context, other approaches have modeled the recommendation problem as a multitask learning of user preferences and item features extracted from reviews, allowing the generation of explanations of model predictions \cite{wang2018explainable, bansal2016ask}. 

Even though these models have taken advantage of information extracted from text reviews, yielding competitive results, none of them have incorporated such information into a KG to enrich the representation of users, items, and their relations. Moreover, using the whole text can often contain redundant and noisy information that can worsen the performance of the model \cite{agarwal2007much}, compared to extracting relevant information in the form of aspects, as proposed in this work. Furthermore, there have been approaches to automatize text content analysis in the field of medicine\cite{carvallo2020automatic, carvallo2019comparing}, along with stress tests\cite{aspillaga2020stress, araujo2021stress, araujo2020adversarial} to evaluate the models performance under different situations. 

\subsection{Aspect-based recommendations} 

Although many aspect-based recommenders exist in the research literature, a few consider incorporating opinions about item aspects into a KG. In general, aspect-based opinions have been used to justify recommendations, including sentiment analysis and text summarization from reviews~\cite{musto2019combining}. 

Several works have exploited aspects to improve the performance of recommender systems. For instance, the work proposed by Chin et al.~\cite{chin2018anr} provides additional knowledge by giving more attention to reviews aspects by means of a neural model with attention mechanisms. Another work done by Musto et al.~\cite{musto2019justifying} proposed to justify recommendations through aspects sentiment analysis on users reviews.  
In terms of combining ratings and text data, a work done by Wu et al.~\cite{wu2015flame} proposed a probabilistic framework that exploits aspect-based opinion mining and collaborative filtering. Moreover, using aspects as features of latent factors~\cite{qiu2016aspect} and neural network models~\cite{chin2018anr, guan2019attentive} have also been investigated. 

Regarding the incorporation of aspect-based opinion information into a KG, TriRank~\cite{he2015trirank} makes use of a tripartite graph representing relationships between users, items, and aspects to generate recommendations. Differently from TriRank, our method combines ratings and aspect-based opinions in a KG to learn richer vector representations for both users and items, and exploits the generated vectors to provide effective personalized recommendations. 

\subsection{Graph-based recommendation explanations} 
The vast majority of works on graph-based recommenders aims to produce path-based interpretations of generated recommendations, although it has been stated that they require domain knowledge~\cite{ma2019jointly}. Fu et al.~\cite{fu2020fairness} made use of paths from users to items, mitigating unfairness problems related to undesired recommendations from inactive users. Xian et al.~\cite{xian2019reinforcement} proposed a reinforcement learning approach for recommendation, where explanations are delivered by a policy-guided path reasoning on a KG. Wang et al.~\cite{wang2019explainable} suggested recommendation explanations by showing items shared features and mining association rules. Lastly, Huan et al.~\cite{huang2019explainable} proposed a method to provide path-based explanation through user-rating interactions incorporating dynamic user preferences. 

Our recommendation method brings entities and relationships in the same latent space, and presents embedding-based explanations, which have been proven to be effective~\cite{ma2019jointly}. Furthermore, we propose a visualization to give more information at the aspect opinion level to final users using the embeddings representations learned from an input KG. 

\section{Proposed method}

In this section, we first introduce our KG model, which incorporates ratings and aspect-based opinions (subsection 3.1). Next, we explain the graph embedding learning process conducted over our KG model (subsection 3.2). 

\subsection{Rating and aspect opinion knowledge graph}

\begin{figure}[t]
    \centering
    \includegraphics[scale=0.5,trim={0mm 7mm 0mm 7mm}]{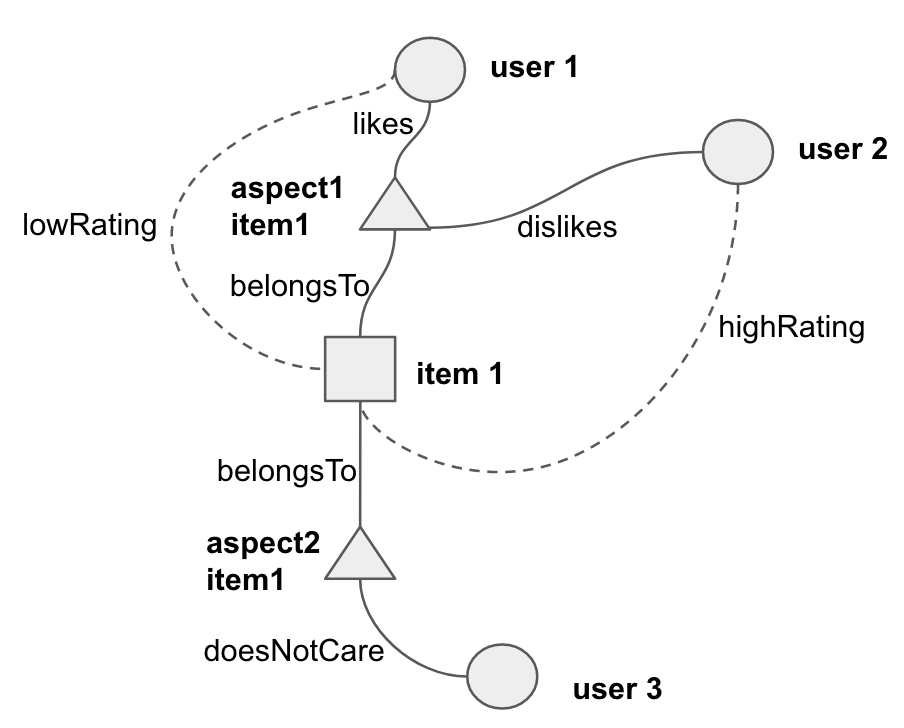}
    \caption{Example of ratings and aspect-based opinions graph. Users may \textit{like}, \textit{dislike} or \textit{doesNotCare} about item aspects, or may assign \textit{highRating} or \textit{lowRating} values to items. An aspect \textit{belongsTo} an item. Circle, triangle and square nodes represent users, aspects and items, respectively. Dotted lines correspond to rating relations, while solid lines are aspect-based opinion relations.}
    \label{graph_fig}
\end{figure}

Our KG representation model is depicted in Figure \ref{graph_fig}. Let $G = (N, R, E)$ be a graph where \textit{N} are nodes representing users, items and item aspects; \textit{R} is the set of possible relations between such entities: \textit{like}, \textit{doesNotCare} and \textit{dislike} for user-aspect relations, \textit{belongsTo} for aspect-item relations, and \textit{highRating} and \textit{lowRating} for user-to-item relations; and \textit{E} represents the graph edges, whose elements are semantic triples $<s,r,d>$ --source, relation and destination--, where $s,d \in N$ and $r \in R$.  
More specifically, to represent ratings given by users to items as triples $<s,r,d>$, we use two types of relations: \textit{highRating} and \textit{lowRating}. Considering a 5-point Likert scale, if a rating is lower than or equal to 3, then it is transformed into a \textit{lowRating} relation; otherwise, it is represented as \textit{highRating} relation. An aspect-based opinion, by contrast, is represented as two triples $<s,r,d>$: one that relates the user and the aspect, and other that relates the aspect and the item. The former relation can be a \textit{likes}, \textit{doesNotCare} or \textit{dislikes} relation depending on the polarity (positive, neutral or negative) of the opinion given to the aspect. This polarity is computed through the opinion mining technique proposed by Hern\'andez-Rubio et al.~\cite{hernandez2019comparative}. If the extracted polarity is $0$, it means the user \textit{doesnotCare} about the aspect for that particular item; if the polarity is greater than $0$, the user \textit{likes} the item's aspect, and finally, if the polarity is lower than $0$, the user \textit{dislikes} the item's aspect. The latter relation is always \textit{belongsTo}, stating that the aspect is associated to the item. 
An illustrative KG example is shown in Figure \ref{graph_fig}, where user1 likes aspect1, which belongs to item1, despite the fact that the item receives a low rating from the user; user2 dislikes the above aspect of item1, but assigns a high rating to the item; and user3 does not care about aspect2 of item1, and does not rate the item.

\subsection{Knowledge graph embedding-based recommendations}

\begin{table}[t]
    \caption{Notation used in the KG embedding learning process.}
    \centering
    \begin{tabular}{c|l}
    \hline 
    Symbol & Description \\    
    \hline
        $\theta_s$ & learnable parameters for source node. \\
        $\theta_d$ & learnable parameters for destination node. \\
        $\theta_r^{(i)}$ & learnable parameters for relations types $r^{(i)}$ \\
        $sim(.)$ & similarity function. \\
        $g(.)$ & transformation function. \\
        $f(.)$ & fitness function that combines similarity ($sim(.)$) and transformation function ($g(.)$). \\
        $f(e)$ & fitness function over original training edges. \\
        $f(e')$ & fitness function over negative sampled edges. \\
        $\lambda$ & margin meta-parameter. \\
        $Re(.)$ & real part of complex vector operation. \\
        $conj(x)$ & conjugate of vector x. \\
        $<\theta_a,\theta_b,\theta_c>$ & trilinear product between embeddings a,b and c. \\
        $\theta_a \odot \theta_b$ & element-wise multiplication between embeddings a and b.\\
        \hline 
    \end{tabular}
    \label{tab:notation}
\end{table}

On the proposed KG, we used PyTorch BigGraph (PBG), a graph embedding learning framework released by Facebook AI~\cite{lerer2019pytorch} for training embeddings on huge graphs. We use this framework since it supports including aspect and rating interactions in a graph-based recommendation system, allowing us to train embeddings for each entity and relations in the same latent space efficiently due to parallelization of the learning process into multiple partitions of the graph. 

To explain the PBG representation learning process, we define $E$ as a set of edges represented as triples $<s,r^{(i)},d>$ that connect source (\textit{s}) and destination (\textit{d}) nodes trough one or many types of relations ($r^{(i)}$). As shown in the previous section, source nodes can be users, items, and aspects depending on type of relations, where \textit{highRating/lowRating} relations directly connect users and items, \textit{like/dislike} relations connect users to aspects, and \textit{belongsTo} relation connects aspects to items (for more details on notation see Table \ref{tab:notation}). 

For each triple, the PBG model have parameters $(\theta_s, \theta_r^{(i)}, \theta_d)$ representing source nodes ($s$), relations ($r^{(i)}$), and destination nodes ($d$), and a fitness function $f$ whose purpose is to reach a maximum in the cases where the triple $e$ belongs to $E$, and to minimize cases when the edge does not exists in the graph. 

To achieve this, $f$ encodes a latent representation learned for $s$, $d$ through different types of relations expressed as $r^{(i)}$, the function is the following:

$$f(\theta_s, \theta_r^{(i)}, \theta_d) = sim(g_s(\theta_s, \theta_r^{(i)}), g_s(\theta_d, \theta_r^{(i)})$$

\noindent
where the transformation function $g$ encodes the relation between nodes and relations $r^{(i)}$ vectors, and the similarity function ($sim$) computes a similarity score between both transformed vectors. 

PBG combines similarity ($sim$) and transformation ($g$) functions to learn embeddings for source nodes, destination nodes, and relations. This combination of functions varies depending on the type of relations being modeled by minimizing the following triple loss function:

$$L = \sum_{e \in E} \sum_{e' \in S'_e} max(f(e) - f(e') +\lambda , 0)$$

\noindent
where $S'_e$ is a set of edges obtained using negative sampling, which samples half of them from corrupting one node from training edges and the other half from triples randomly created, and $\lambda$ is a margin hyperparameter. 

In this case, as we have multiple relations and different types of nodes, we use $Re(<\theta_a, conj(\theta_b) >)$ as similarity function. \textit{Re(.)} denotes the real part of a complex vector, $conj(.)$ is the complex conjugate of the vector and $<.,.,.,>$ denotes the trilinear product.

Then, as transformation function, we use element-wise multiplication ($\odot$) between the resulting transformed vector and the relation vector $\theta_r$. Given this setup, the function \textit{f} is defined as follows:

$$f(\theta_s, \theta_r^{(i)}, \theta_d) = Re(< \theta_s \odot \theta_r^{(i)} , conj(\theta_d \odot \theta_r^{(i)}) >) $$

\noindent
which corresponds to a complEX \cite{trouillon2016complex} learning representation strategy, since we empirically obtained that it was the best combination of transformation and similarity functions, in terms of recommendation performance compared to transE~\cite{bordes2013translating}, transR~\cite{lin2015learning}, RESCAL~\cite{nickel2011three} and Distmult~\cite{yang2014embedding}. 

Finally, to generate the recommendations, as user and item embeddings are in the same latent space, we look for the item embeddings closest to the target user's embedding using cosine similarity. 





\section{Experiments}

In this section, we report the conducted experiments, describing the used datasets (subsection 4.1), presenting the evaluated recommendation methods (subsection 4.2), and discussing the achieved results (subsection 4.3).

\subsection{Datasets}
To evaluate the proposed method, we used the well-known Amazon Reviews dataset\footnote{https://jmcauley.ucsd.edu/data/amazon/} (AMZ) with reviews and ratings for Movies \& TV (MVT), Videogames (VGM), and Cellphones (CPH). For this dataset, we considered only users with more than $10$ ratings. We also experimented with other domains --Restaurants (RST), Hotels (HTL), and Beauty\&Spa (BTY)-- using the Yelp Challenge dataset\footnote{https://www.yelp.com/dataset} (YLP), which, as the Amazon dataset, contains both ratings and reviews. 

Additionally, enriching the above datasets, we used the aspect opinions dataset\footnote{http://ir.ii.uam.es/aspects/} provided by Hern\'andez et al.~\cite{hernandez2019comparative}, which complements the ratings with aspect-based opinions (i.e., [user, item, aspect, opinion polarity] tuples) extracted from the Amazon and Yelp reviews.

Table \ref{table-datasets} shows statistics of the datasets. It can be seen that the number of ratings varies from 25 thousand to more than one million and three thousand interactions. It can also be observed that the Yelp dataset for HTL and BTY domains has fewer interactions compared to others. Regarding aspect-based opinions, the number of relations varies from 13 thousand to more than 2 millions for the MVT domain. Furthermore, in general, users are more active for rating interactions than aspect opinions, except for the MVT domain.

\begin{table}[h]
\small
\caption{Datasets statistics}
\vspace{-3mm}
\label{table-datasets}
\centering
{
\begin{tabular}{ccccccccc} 
    \toprule
    {Dataset} & {Domain} & {Users} & {Items} & {Ratings}  & {Aspect} & {Rating} \\
    &&&&&opinions&sparsity
    \\ \hline
    
    \multirow{3}{*}{\makecell{AMZ}} & {MVT} & 40,058 & 131,638 & 1,349,351 & 2,132,927 & $2.4 \cdot 10^{-4}$   \\
    {} & {VGM} & 6,686 & 25,795 & 575,136 & 209,755 & $8.3 \cdot 10^{-4}$    \\ 
    {} & {CPH} & 7,237 & 47,183 & 464,684 & 85,930 &  $3.4 \cdot 10^{-4}$  \\ 
    
    \hline
    
    \multirow{3}{*}{\makecell{YLP}} & {RST} & 36,471 & 4,503 & 791,865 & 547,892 & $9.6 \cdot 10^{-4}$  \\
    {} & {HTL} & 4,148 &  284 & 25,170 & 16,935 & $4.2 \cdot 10^{-3}$ \\ 
    {} & {BTY} & 4,270 & 764 & 27,885 & 13,217 &  $1.7 \cdot 10^{-3}$  \\
    \bottomrule
    \end{tabular}}
\end{table}

\subsection{Evaluated methods}
We report experimental results from the following methods:

\begin{itemize}
    
    \item \textbf{Non-personalized recommendation methods:} random rating (RDM) and most popular item (POP) recommendation methods. 
    
    \item \textbf{Matrix Factorization optimized with alternate least squares (MF):} a traditional baseline; a matrix factorization model optimized with alternate least-squares~\cite{koren2009matrix} using 200 latent factors, which yielded better results than other traditional baselines.
    
    \item \textbf{Graph-based recommendation baseline (KGAT): } as a graph-based recommendation state-of-the-art, Wang et al.'s KG Attention Network recommendation method~\cite{wang2019kgat} . 
    
    \item \textbf{Graph embeddings from ratings (GER):} a version of our proposed method that considers only rating-based relations between users and items.
    
    \item \textbf{Graph embeddings from aspect-based opinions (GEA):} a version of our proposed method that considers \textit{likes}, \textit{dislikes}, and \textit{doesNotCare} relations between users and item aspects, and \textit{belongsTo} relation between item aspects and items. 
    
    \item \textbf{Graph embeddings from ratings and aspect-based opinions (GERA):} our proposed method  considering both rating- and aspect opinion-based relations.
    
\end{itemize}

We note that content-based filtering, user-based collaborative filtering~\cite{schafer2007collaborative}, item-based collaborative filtering~\cite{schafer2007collaborative}, matrix factorization using alternated least-squares~\cite{koren2009matrix}, and matrix factorization using Bayesian personalized ranking~\cite{rendle2012bpr} were also evaluated, performing worse. 

\subsection{Achieved results}

In an experimental setup phase, we performed a grid search testing a range of parameters, obtaining best performance results for embeddings of dimension 400, a learning rate of 0.01, 5 epochs, and a hinge loss margin of 0.1. 

For the evaluation, we used the OrdRec methodology~\cite{koren2013collaborative}, where a recommendation method is evaluated on the test ratings at the user level. We employed a 5-fold cross-validation strategy, and computed the F1@k (k = 10,20,30) metric. Other metrics such as nDCG@k (k = 10,20,30), recall@k (k = 10,20,30), and precision@k (k = 10,20,30) were also calculated, but they did not give additional insights to the analysis we present next. 

The results in Tables \ref{amazon_results_table} and \ref{yelp_results_table} indicate that in most of the domains, our method (GERA) outperforms the baselines in terms of F1@k (10,20, 30). For a top-N recommendation task, GERA outperforms state-of-the-art traditional and graph-based recommendation methods. However, for the CPH domain, in terms of F1@10, MF achieves the highest value (0.915), whereas, for F1@20 and F1@30, the best-performing method was the aspect-based opinion graph (GEA) version of our proposal, with values of 0.935 and 0.939, respectively. The worse results of GERA in the CPH domain can be explained because, as shown in Table \ref{table-datasets} this domain does not have enough aspect opinions to improve the results by combining them with ratings in a KG. 

Besides, when comparing graph-based models that use only ratings (GER) with the ones that consider only aspects (GEA), we observe that in the CPH, RST, HTL and BTY domains, the GEA model performs better than GER. This can be explained because, in addition to the fact that these domains have a considerable amount of aspect interactions (cf. \ref{table-datasets}). 

Furthermore, the model that uses only ratings (GER) performs better on the MVT and VGM domains than those that use only aspects (GEA). This behavior can be due to the vast amount of rating interactions in these particular domains (cf. \ref{table-datasets}). However, an interesting phenomenon is that when combining both models into one that considers both rating and aspect opinion interactions (GERA), we have better results than GEA and GER separately.

\begin{table}[t]
        \small
        \captionof{table}{Evaluation results on the Amazon dataset. Values in bold are the best ones for each domain and metric. * indicates statistical significance best result by multiple t-test using Bonferroni correction.}
        
        \label{amazon_results_table}
        
        \begin{tabular}{cccccc}
        \toprule
         Dataset & Domain & Model & F1@10 & F1@20 & F1@30  \\ \hline
    
         \multirow{6}{*}{}& {}  & {RDM}  & {.801}  & {.812} & {.813}   \\
         && POP & .033 & .031 & .031    \\
         && MF & .905 & .914 & .917   \\
         & MVT &KGAT & .924 & .938 & .942 \\
         &&GER & .921 & .936 & .940   \\
         &&GEA & .915 & .932 & .936   \\
         &&GERA & \textbf{.928*} & \textbf{.941*} & \textbf{.945*}  \\

        \cline{2-6}
        
         && RDM & .800 & .807 & .806  \\
         && POP & .045 & .045 & .045    \\
         && MF & .909 & .914 & .915 \\ 
         AMZ & VGM &KGAT & .791  & .903 & .920  \\
         && GER & .791 & .901 & .920  \\
         &&GEA  & .781 & .898 & .918 \\
         &&GERA & \textbf{.934*} & \textbf{.943*} & \textbf{.945*}      \\

        \cline{2-6}
        
        && RDM  & .819 & .819 & .823   \\
        && POP & .043  & .043 & .043   \\
        && MF & \textbf{.915*} & .916 & .917             \\ 
        & CPH &KGAT & .888 & .929  & .933 \\
        && GER & .890 & .930 & .934 \\
        && GEA & .898 & \textbf{.935*} & \textbf{.939*} \\
        && GERA & .880 & .928 & .933   \\

        \bottomrule
        
        \end{tabular}
\vspace{3mm}
        \small
        \captionof{table}{Evaluation results on the Yelp dataset. Values in bold are the best ones for each domain and metric. * indicates statistical significance best result by multiple t-test using Bonferroni correction.}
        \label{yelp_results_table}
        \begin{tabular}{cccccc}
        
        \toprule
        
        Dataset & Domain & Model & F1@10 & F1@20 & F1@30  \\ \hline
        
        \multirow{3}{*}{} &  &  RDM & .647 & .646 & .650 
          \\
         && POP & .107 & .107 & .107   \\
         && MF & .827 & .829 & .830    \\ 
         & RST &KGAT & .913 & .932 & .938 \\
         && GER & .914  & .931  & .936 \\
         && GEA & .927 & .940 & .944 \\
         &&GERA & \textbf{.933*} & \textbf{.944*} & \textbf{.947*}             \\ 
         
         \cline{2-6}
        
         && RDM & .592 & .589 & .571    \\ 
         && POP &.468 & .468 & .468   \\
         && MF & .816 & .816 & .816    \\
         YLP & HTL &KGAT & .986 & .986 & .986\\
         &&GER & .983 & .983 & .983 \\
         &&GEA & .985 & .985 & .985 \\
         &&GERA & \textbf{.987*} & \textbf{.987*} & \textbf{.987*}    \\

        \cline{2-6}
        
         && RDM & .650 & .634 & .622     \\
         && POP & .263 & .263 & .263   \\
         && MF & .846 & .848 & .850    \\
         & BTY &KGAT & .984 & .984 & .984\\
         &&GER& .983 & .984 & .984 \\
         &&GEA& .983 & .983 & .983  \\
         &&GERA & \textbf{.987*} & \textbf{.988*} & \textbf{.988*}    \\ 
         
        \bottomrule
        \end{tabular}
\end{table}

\section{Recommendation explanations}
In this section, we present a simple technique to generate explanations about recommendations provided by our method. To obtain interpretable recommendations at aspect level, we first recommend 30 (a chosen cut-off) items whose embeddings are closer to the target user's embedding. From them, we search for the similar users, and gather the opinions these users made about aspects from the recommended items. 
We first present our proposed explainable recommendation dashboard (subsection 5.1), and then we report statistics associated to explanations on each of the addressed datasets and domains (subsection 5.2).

\subsection{Explainable dashboard}

In order to provide a concrete usage scenario about how our model can be leveraged to produce a rich interactive user experience, we designed a visualization dashboard. We followed Munzer nested visualization design framework \cite{munzner2014visualization} in order to reduce the design space and prevent ineffective visualizations. We then start by answering the three questions:
\textit{What} (input data for the dashboard), \textit{Why} (visual tasks) and \textit{How} (the actual visualization and interactive encodings).

\textbf{What}. The input data for the dashboard are: A given user $u$, the list of movies recommended with their corresponding score (score is data of type quantitative), 2-dimensional UMAP \cite{mcinnes2018umap} embeddings for each user in the dataset, set of aspects liked and disliked by each user on each movie reviewed (with the corresponding frequency per user), record of ratings per user, review text per user with aspects identified, pre-computed.

\textbf{Why}. The visual tasks in Munzner framework are presented in the the form of pairs \{\textit{action}, \textit{target}\}. We identify the following visual tasks for designing the dashboard: 
\begin{itemize}
    \item \{\textit{present}, \textit{recommendation}\}: a visual element must let users see the list of recommended items and trigger actions if the user selects an item.
    \item \{\textit{browse}, \textit{neighborhood}\}: the user should be able to select an arbitrary group users in order to conduct other visual tasks.
    \item \{\textit{compare}, \textit{aspect distribution}\}: The user must the able to compare her own distribution of aspects (liked and disliked) with the distribution of aspects from other users arbitrarily selected.
    \item \{\textit{locate}, \textit{outliers}\}: The user should be able to identify outlier users or neighborhoods in order to help them interpret her own recommendations.     
    \item \{\textit{identify}, \textit{features}\}: The user should be able to identify features (aspects in our case) in the textual description that she and other people have written.
    \item \{\textit{identify}, \textit{aspect distribution}\}: the user must be able to visualize her own liked and disliked aspects.
    \item \{\textit{identify}, \textit{rating distribution}\}: the user must be able to visualize the rating distribution of arbitrary people selected.
    \item \{\textit{identify}, \textit{extremes}\}: The user should be able to identify extreme users in terms of aspects they care about reviewing.
\end{itemize}

\textbf{How}. Once the visual tasks are identified, it is time to choose the actual visual and interactive elements. The first decision we made here is actually designing several views, where in each view we can see a different visual element which will support the visual tasks identified in the previous step. In terms of interactive design, all these views will be connected in order to update the plots once the user selects certain elements in the dashboard. Figure \ref{explanation_dashboard} shows an illustrative example where our proposed visualization dashboard presents generated aspect-level explainable recommendations consisting of:

\begin{figure}[t]
\vspace{-5mm}
  \makebox[\textwidth]{\includegraphics[width=15cm]{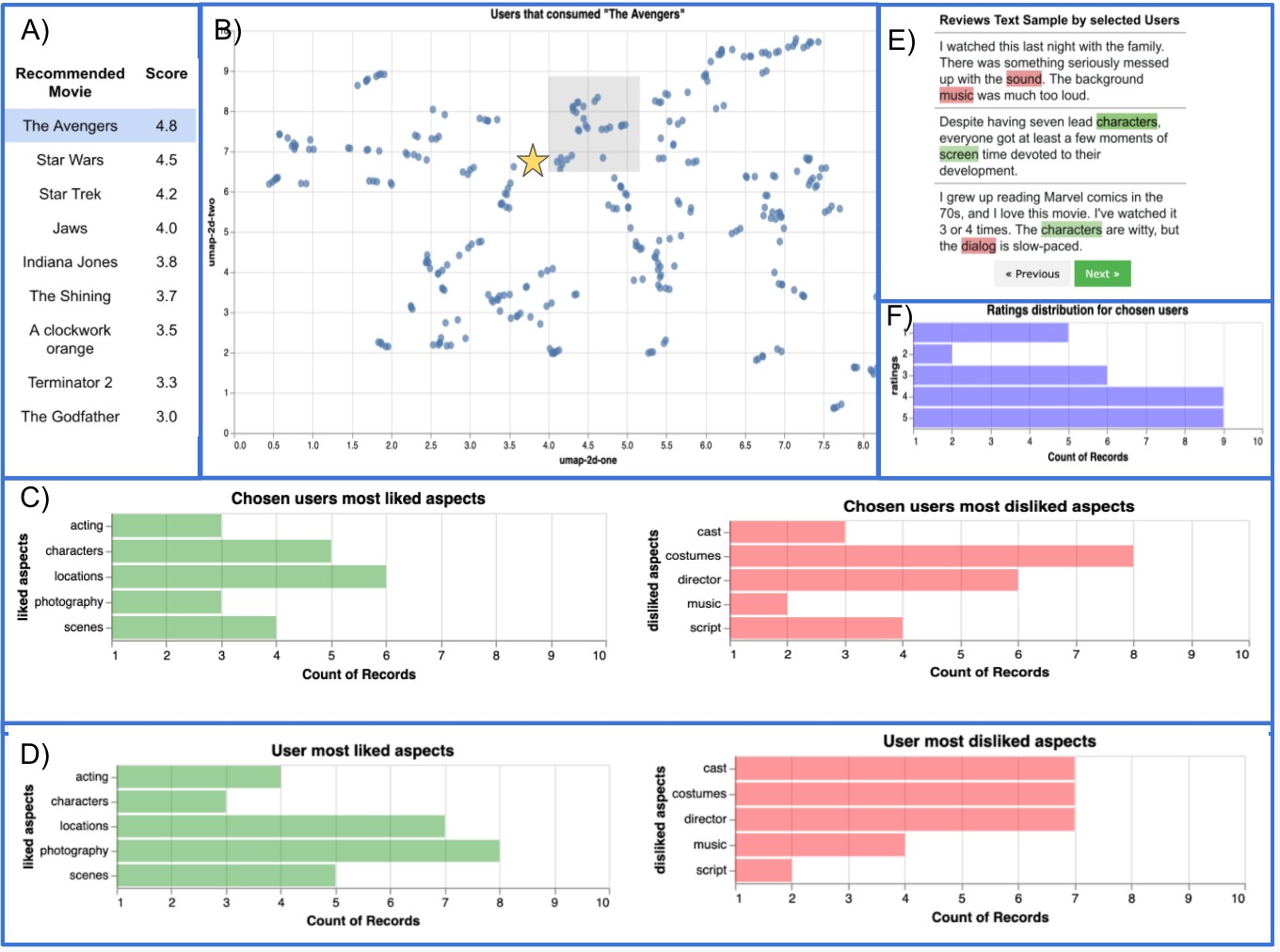}}
  \caption{Visual dashboard for interactive model exploration. \textbf{(A)} recommendations where user can select one movie \textit{M}, \textbf{(B)} this panel shows all the users who watched movie \textit{M}, each user represented by 2D UMAP embedding, the active user is presented by a star. \textbf{(C)} distribution of liked and disliked aspects for users selected in previous (B) step, \textbf{(D)} distribution of liked items and disliked items for active user,  \textbf{(E)} sampled review texts by people selected in (B) with highlighted words corresponding to aspects, and \textbf{(F)} rating distribution of users chosen in (B).}
  
\label{explanation_dashboard}
\end{figure}

\begin{enumerate}[label=\Alph*)]
    \item Recommended items and associated scores given by our recommendation method. The user can select one of the recommended items to display other associated visualizations.
    \item 2D chart of users that interacted with a chosen item. The dimensionality of the graph-learned user embeddings is reduced by UMAP. In the dashboard interface, one can select a neighborhood, where the current user is symbolized with a star.
    \item Barplots for most liked and most disliked aspects from the chosen neighborhood given a selected recommended item.
    \item Barplots for most liked and most disliked aspects from the current user.
    \item Sampled reviews text for a selected item, indicating the liked aspects in green, and the disliked aspects in red. It is worth mentioning that the model captures important aspects and aspect synonyms in the reviews. Also, it is possible to explore an item's reviews with a previous and next button. 
    \item Distribution of ratings given by the subset of selected users to the chosen item. 
\end{enumerate}

\textbf{Use case}. Following the example from Figure \ref{explanation_dashboard}, an active user was recommended ten movies, of which he chooses ``The Avengers'' (see Figure \ref{explanation_dashboard}A). By choosing this movie, all the users who have commented on the movie are displayed in a 2D plane (Figure \ref{explanation_dashboard}B). The active user can then select a subgroup of closest neighbors, and obtain the aspects they liked and disliked the most. In the example, the most-liked aspect of the chosen neighborhood is ``locations,'' with six liked interactions, and the most unliked aspect is ``costumes,'' with eight dislike interactions (Figure \ref{explanation_dashboard}C). Then, the active user can be compared with others based on their most liked and disliked aspect distributions (Figure \ref{explanation_dashboard}D); in the example, the most liked aspect of the active user is ``photography,'' and the most disliked aspects are ``cast,'' ``costume'' and ``director''.
Exploring the distribution of ratings (Figure \ref{explanation_dashboard}F) of the chosen users for the movie ``Avengers,'' nine users gave ratings 4 and 5. The other rating values are distributed among other users, showing that the recommendation score is similar to the nearest users. 
In the dashboard, one can also explore examples of reviews with the positive aspect-based opinions highlighted in green and the negatives aspect-based opinions highlighted in red (Figure \ref{explanation_dashboard}E).

\subsection{Statistics of recommendation explanations}

In Table \ref{table-intepretability}, we report \textit{Coverage} --percentage of items with aspect opinion-based explanations--, \textit{lk/other} --the proportion of \textit{likes} over \textit{dislikes} and \textit{doesNotCare}) relations--, \textit{\#aspects} --the average number of unique aspects--, and \textit{asp/item} --the average number of aspects per item.

\begin{table}[H]
\caption{Statistics of the recommendation explanation technique for each domain. Results in bold are the best ones for each metric.}
\setlength{\tabcolsep}{0.45em} 
{\renewcommand{\arraystretch}{0.95}
\label{table-intepretability}
\centering
\vspace{-3mm}
{
\small
\begin{tabular}{ccccccccc} \toprule
    {Dataset} & {Domain} & {Coverage} & {lk/other} & {\#aspects} & {asp/item}    \\ \hline
    
    \multirow{3}{*}{\makecell{AMZ}} & {MVT} & 47.90\% & .663 & 10.30 & .716   \\
    {} & {VGM} & \textbf{90.23}\% & .584  & 18.52  & .684     \\ 
    {} & {CPH} & 70.38\% & .592 & 12.27  & .581  \\ 
    
    \hline
    
    \multirow{3}{*}{\makecell{YLP}} & {RST} & 11.11\% & .508 & 6.22 & 1.87   \\
    {} & {HTL} & 21.64\% & \textbf{.796} & \textbf{22.34} & \textbf{3.44} \\ 
    {} & {BTY} & 20.02\% & .653 & 17.55 & 2.88  \\ 
    
    \bottomrule
    
    \end{tabular}}}
\end{table}

\vspace{-2mm}    

These results indicate that the proposed method has better coverage in the VGM domain with a 90.23\% value; this shows that of the 30 recommended items, 27 have an interpretation at the aspect level. Regarding the balance of likes with respect to other interactions, the proposed method in the HTL domain has the highest ratio (0.796). This means that, in general, the aspects of hotels are valued higher than in other domains. The method explanations in that domain cover a greater variety of aspects, with an average of 22.34 unique aspects per recommendation. More evidence of this is given by the high number of aspects per item (3.44). However, despite the diversity in explanations, the method in the HTL domain has the third worst coverage, followed by RST and BTY. Comparing the results on the two datasets, it can be seen that in terms of coverage, in general, AMZ surpasses YLP.

An interesting behavior to discuss is if there is a relationship between models' interpretability in terms of coverage and their recommendation performance, where both seem to have a negative interdependence. For example, although the proposed model in the RST domain has the best performance results, it has the lowest coverage (11.11\%), showing that there seems to be a trade-off between explainability and models' performance. A similar response is observed for HTL and BTY domains. 

\section{Conclusions} 
In this paper, we have empirically shown that latent representations of users and items extracted from a knowledge graph by means of rating-based embeddings can be exploited to effectively outperform the performance of graph-based and traditional recommendation approaches. Moreover, we have shown that richer embeddings integrating ratings and aspect-based opinions in the knowledge graph allow surpassing a state-of-the-art neural network recommendation model. 

In addition to performance improvements, an advantage of the proposed method is the interpretabilty of its recommendations, since  users, items, and aspects are modeled in a common latent space. Differently from previous approaches that generate explanations at rating level or according to graph paths that require domain knowledge, our method provides aspect-level explanations extracted from item reviews.

In the paper, we have provided an example of an explanatory visualization dashboard that presents generated aspect-level explanations for recommendations, showing the potential that the graph-based method can have in terms of interpretability. A user study aimed to evaluate the potential benefits (e.g., system transparency, and user trust and satisfaction) of the recommendation explanations is left as future work.

Our work has shown that to improve state-of-the-art graph-based recommendations, it is not always necessary to change the recommendation algorithm, but preferably seek new ways of incorporating existing user preference information like ratings and aspect-based opinions into a reference knowledge graph. Furthermore, other information that will be relevant to add to the recommendation are the embeddings learned from aspects since, in this work, we only use those of users and items.

Insightful results were achieved when testing two variants of the proposed graph-based recommendation method: one that considers only ratings (GER) and other that only uses aspects (GEA). We observe that GEA has better recommendation performance on domains that have a large amount of aspect opinion interactions. In contrast, when the number of rating interactions was relatively high, GER outperformed GEA. Interestingly, when combining rating and aspect opinion information in a graph model (GERA), we significantly outperformed both GEA and GER models separately, among other baselines. 

As future work, a user study will be performed to demonstrate whether the proposed dashboard that incorporates explanations on aspects is helpful for end-users.

\bibliographystyle{ACM-Reference-Format}
\bibliography{sample-base}

\end{document}